\documentclass[12pt,leqno,a4paper]{article}
\usepackage{amsmath,amsthm}
\usepackage[latin1]{inputenc}
\usepackage[T1]{fontenc}
\usepackage[english]{babel}

\newtheorem{theorem}{Theorem}[section]
\newtheorem{proposition}[theorem]{Proposition}

\newtheorem{corollary}[theorem]{Corollary}
\newtheorem{definition}[theorem]{Definition}

\newtheorem{remark}[theorem]{Remark}


\newcommand*\proofnamestyle{\itshape}

\DeclareMathOperator{\tr}{Tr}

\begin{document}

    \title{Extensions of Lieb's concavity theorem}
      \author{Frank Hansen}
      \date{November 28, 2005\\
      {\tiny Revised July 14, 2006}}
      \maketitle

      \begin{abstract}
      The operator function $ (A,B)\to\tr f(A,B)(K^*)K, $ defined in pairs of
      bounded self-adjoint operators in the domain of a function $ f $ of two real variables, is convex for
      every Hilbert Schmidt operator $ K, $ if and only if $ f $ is operator convex.
      We obtain, as a special case, a new proof of Lieb's concavity theorem for the function
      $ (A,B)\to\tr A^pK^*B^{q}K, $ where $ p $ and $ q $ are non-negative numbers with sum $ p+q\le 1. $
      In addition, we prove concavity of the operator function
      \[
      (A,B)\to\tr\left[\frac{A}{A+\mu_1}K^*\frac{B}{B+\mu_2}K\right]
      \]
      in its natural domain $ D_2(\mu_1,\mu_2), $ cf. Definition~\ref{definition: domain of concavity}.
      \end{abstract}

    \section{Introduction}

    Let $ f:D\to{\mathbf R} $ be a function of two variables defined in a set $ D\subseteq{\mathbf R}^2, $
    and let $ M_{n\times m} $ denote the set of complex $ n\times m $
    matrices (with the abbreviation $ M_n $ for $ M_{n\times n}). $
    We say that two Hermitian matrices $ (A,B)\in M_n\times M_m $ are in the domain of $ f, $ if the
    product $ \sigma(A)\times\sigma(B) $ of the spectra is included in $ D. $ We shall consider two
    different but related notions of matrix functions associated with $ f. $

    \subsection{The functional calculus}

     Following Kor{\'a}nyi \cite{kn:koranyi:1961}, we introduce the functional calculus
    \begin{gather}
    f(A,B)=\sum_{i=1}^p\sum_{j=1}^q f(\lambda_i,\mu_j) P_i \otimes Q_j
    \end{gather}
    for functions $ f $ of two variables, where
    \begin{gather}\label{spectral decompositions}
    A=\sum_{i=1}^p \lambda_iP_i\quad\mbox{and}\quad B=\sum_{j=1}^q \mu_j Q_j
    \end{gather}
    are the spectral decompositions of $ A $ and $ B. $ If $ f $ can be written as a product
    $ f(t,s)=g(t)h(s) $ of two functions each depending only on one variable then $ f(A,B)=g(A)\otimes h(B). $
    We say that $ f $ is matrix convex of order $ (n,m), $ if $ D $ is convex and
    \[
    f(\lambda A+(1-\lambda)B,\lambda C+(1-\lambda)D)\le\lambda
    f(A,C)+(1-\lambda)f(B,D)
    \]
    for all pairs of Hermitian matrices $ (A,C), (B,D)\in M_n\times M_m $ in the
    domain of $ f $ and $ \lambda\in[0,1]. $ Note that $ (\lambda A+(1-\lambda)B,\lambda
    C+(1-\lambda)D) $ automatically is in the domain of $ f. $

    This type of functional calculus may for continuous functions be
    extended to bounded, linear and self-adjoint operators on a Hilbert space by replacing
    sums with integrals, hence
    \begin{gather}\label{functional calculus for operators}
    f(A,B)=\int f(\lambda,\mu)\, dE_A(\lambda) \otimes dE_B(\mu),
    \end{gather}
    where $ E_A\otimes E_B $ is the product measure constructed from the two spectral measures $ E_A $ and $ E_B. $
    It is well-defined on products of Borel sets in $ \mathbf R $ since $ E_A\otimes 1 $ and $ 1\otimes E_B $
    commute, and it may be extended to Borel sets in $ \mathbf R^2. $ The support of
    the measure is contained in $ \sigma(A)\times\sigma(B). $

    The function $ f $ is said to be operator
    convex, if $ D $ is convex and the operator function $ (A,B)\to f(A,B) $ is convex in pairs of
    operators in the domain of $ f. $ It is not difficult to establish that
    $ f $ is operator convex, if an only if it is matrix convex of all
    orders. The proof follows a suggestion by L{\"o}wner (for operator monotone functions) as reported by Bendat and
    Sherman~\cite[Lemma 2.2]{kn:bendat:1955}
    and can easily be adapted to the present situation. Note finally that this type of functional calculus
    may be generalized to functions of $ k $ variables, together with the notion of operator convexity
    or matrix convexity of a fixed order $ (n_1,\dots,n_k). $

    \subsection{The variant functional calculus}

    We may also define an endomorphism  $ K\to f(A,B)(K) $ of $ M_{n\times m} $ by setting
    \begin{gather}\label{reduced functional calculus for operators}
    f(A,B)(K)=\sum_{i=1}^p\sum_{j=1}^q f(\lambda_i,\mu_j) P_i K Q_j
    \end{gather}
    for each $ K\in M_{n\times m}. $ If $ f $ can be written as a product
    $ f(t,s)=g(t)h(s) $ of two functions each depending only on one variable then $ f(A,B)(K)=g(A)K h(B). $
    This type of functional calculus is difficult to extend to bounded linear operators
    on a Hilbert space $ H, $ since there is no obvious way of constructing a measure on
    $ H $ from the two spectral measures $ E_A $ and $ E_B. $
    These questions ''were extensively investigated
    by Birman and Solomyak~\cite{kn:birman:1972,kn:birman:1973}
    within the very general scope of their theory of double operator
    integrals'', and it is only possible to extend the type of functional calculus in
    (\ref{reduced functional calculus for operators})
    to bounded linear operators for a special class of functions, cf. also \cite{kn:hiai:2003:1}.
    The variant functional calculus is in the literature sometimes expressed in terms of ''super operators''
    acting on $ M_{n\times m} $ by setting
    \[
    f(A,B)(K)=f(L_A,R_B)K,
    \]
    where $ L_A $ and $ R_B $ are commuting left and right multiplication operators (by $ A $ and $ B). $

    \subsection{Convexity statements}

    The two types of functional calculus are connected by the following construction.
    Let $ H_1 $ and $ H_2 $ be Hilbert spaces of finite dimensions $ n_1 $ and $ n_2 $
    equipped with fixed ortho\-normal bases
    $ (e^1_1,\dots,e^1_{n_1}) $ and $ (e^2_1,\dots,e^2_{n_2}). $ Let furthermore
    \[
    \{e_{ij}\}_{i=1,\dots,n_1;\,j=1,\dots,n_2}
    \]
    be the system of matrix units in $ B(H_2,H_1) $ such that
    \[
    e_{ij}e^2_m=\delta_{jm}e^1_i\qquad j,m=1,\dots,n_2;\, i=1,\dots,n_1.
    \]
    Let $ \bar H_2 $ denote the Hilbert space conjugate\footnote{
    This means that $ H_2 $ and $ \bar H_2 $ are identical as complex vector spaces, but the
    inner products are conjugate to each other.}
    to $ H_2 $ and consider the linear bijection $ \Phi\colon H_1\otimes \bar H_2\to B(H_2, H_1) $ such that
    \[
    \Phi(e^1_i\otimes e^2_j)=e_{ij}\qquad i=1,\dots,n_1;\, j=1,\dots,n_2.
    \]
    It is not difficult to establish that $ \Phi $ is unitary and that
    \begin{equation}
    \Phi(f(A,B)\varphi)=f(A,B)(\Phi(\varphi)),
    \end{equation}
    hence
    \begin{equation}
    (f(A,B)\varphi\mid\varphi)_{H_1\otimes\bar H_2}=\tr\left(f(A,B)(\Phi(\varphi))\Phi(\varphi)^*\right)
    \end{equation}
    for self-adjoint operators $ (A,B) $ in the domain of $ f $ such that $ A $ is acting on $ H_1 $
    and $ B $ is acting on $ H_2, $ and every vector $ \varphi\in H_1\otimes\bar H_2. $
    We consequently obtain:

    \begin{theorem}\label{main theorem}
    Let $ f:D\to{\mathbf R} $ be a function defined in a convex set $ D\subseteq{\mathbf R}^2. $ The
    matrix function
    \[
    (A,B)\to\tr f(A,B)(K^*)K,
    \]
    defined in pairs of Hermitian matrices $ (A,B)\in M_n\times M_m $ in the domain of $ f, $ is
    convex for all matrices $ K\in M_{m\times n} $ if and only if $ f $ is matrix convex of order $ (n,m). $
    \end{theorem}

    Lieb's concavity theorem states that the mapping
    \[
    (A,B)\to \tr A^pK^*B^q K,
    \]
    defined in pairs of positive definite operators, is concave for arbitrary Hilbert Schmidt operators $ K $
    and non-negative exponents $ p $ and $ q $ with $ p+q\le 1. $
    Let us therefore, for these exponents, consider the function $ f(t,s)=t^{p}s^{q} $ defined in the
    first quadrant. Since
    \[
    \tr f(A,B)(K^*)K=\tr A^pK^*B^q K
    \]
    we realize by Theorem~\ref{main theorem} that Lieb's concavity theorem
    is a reflection of the operator concavity of the function $ f. $
    But Theorem~\ref{main theorem} also sets the scope for the largest possible
    extension of Lieb's theorem, not only for operators but for each class
    of matrices. These distinctions are significant because of the richness
    of the class of matrix convex functions. In a forthcoming paper \cite{kn:hansen:2005:3} we show that
    there to any interval $ I $ different from the real line and to each natural
    number $ n $ exist a function in $ I $ which is matrix convex of order $ n, $
    but not matrix convex of order $ n+1. $

    \section{Some operator concave functions}

    In this section we study some well-known operator concave functions with the aim
    to give truly elementary or otherwise illuminating proofs. The basic tool is
    the geometric mean $ \# $ for positive operators $ A $ and $ B $ introduced
    by Pusz and Woronowicz \cite{kn:pusz:1975,kn:ando:1979,kn:kubo:1980}.
    It is increasing, concave and given by
    \[
    A\,\#\,B=A^{1/2}(A^{-1/2}BA^{-1/2})^{1/2}A^{1/2},
    \]
    if $ A $ is invertible. Note that $ A\,\#\,B=(AB)^{1/2} $ if $ A $ and $ B $ commute.
    The geometric mean $ A\,\#\,B $ may be characterized as the maximum of all self-adjoint $ C $ such that
    the block matrix
    \[
    \begin{pmatrix} A & C\\ C & B\end{pmatrix}
    \]
    is positive semi-definite.
    Adapting the reasoning in \cite[Corollary 2.2]{kn:ando:1979} we obtain:

    \begin{proposition}\label{corollary: product mean of operator concave functions}
    Let $ f $ and $ g $ be non-negative operator concave functions of $ k $ variables defined
    in some convex domain $ D $ in $ {\mathbf R}^k. $ The function
    \[
    F(t_1,\cdots,t_k)=f(t_1,\cdots,t_k)^{1/2}g(t_1,\cdots,t_k)^{1/2}
    \]
    is then also operator concave in the domain $ D. $
    \end{proposition}

    \proof We consider $ k $-tuples  $ (A_1,\dots,A_k) $ and  $ (B_1,\dots,B_k) $ of self-adjoint operators
    in the domain $ D $ and note that
    \[
    F(A_1,\dots,A_k)=f(A_1,\cdots,A_k)\,\#\,g(A_1,\dots,A_k).
    \]
    The statement now follows from the calculation
    \[
    \begin{array}{l}
    \displaystyle F\left(\frac{A_1+B_1}{2},\dots,\frac{A_k+B_k}{2}\right)\\[3ex]
    =\displaystyle f\left(\frac{A_1+B_1}{2},\cdots,\frac{A_k+B_k}{2}\right)\,\#\,
    g\left(\frac{A_1+B_1}{2},\dots,\frac{A_k+B_k}{2}\right)\\[3ex]
    \ge\displaystyle\frac{f(A_1,\dots,A_k)+f(B_1,\dots,B_k)}{2}\,\#\,
    \frac{g(A_1,\dots,A_k)+g(B_1,\dots,B_k)}{2}\\[3ex]
    \ge\displaystyle\frac{f(A_1,\dots,A_k)\,\#\, g(A_1,\cdots,A_k)}{2}+
    \frac{f(B_1,\dots,B_k)\,\#\, g(B_1,\cdots,B_k)}{2}\\[3ex]
    =\displaystyle\frac{F(A_1,\dots,A_k)+F(B_1,\cdots,B_k)}{2},
    \end{array}
    \]
    where we used the concavity of $ f $ and $ g $ and monotonicity of the geometric mean in the first inequality, and the concavity
    of the geometric mean in the second.\qed

    Note that the above proposition may be formulated also for classes of matrix concave functions of a
    fixed order $ (n_1,\dots,n_k). $

    \begin{corollary}\label{corollary: operator concavity of t^p}
    The functions $ (t_1,\dots,t_k)\to t_1^{p_1}\cdots t_k^{p_k} $
    are operator concave in $ \mathbf R^k_+ $ for non-negative exponents $ p_1,\dots,p_k $ with
    sum $ p_1+\cdots+p_k\le 1. $
    \end{corollary}

    \begin{proof}
    Consider the simplex $ S=\{(p_1,\dots,p_k)\mid p_i\ge 0, p_1+\cdots+p_k\le 1\} $ and
    the set of exponents
    \[
    E=\{(p_1,\dots,p_k)\in S\mid t_1^{p_1}\cdots t_k^{p_k}\text{ is operator concave in
    $ \mathbf R_+^k $ }\}.
    \]
    The vertices $ (0,0,\dots,0) $ and
    $ (1,0,\dots,0), $ $ (0,1,\dots,0),\dots,(0,0,\dots,1) $ of the convex polytope $ S $ are in $ E, $ hence
    $ S=\text{conv}(E). $ Since $ E $ is closed and mid-point convex by
    Proposition~\ref{corollary: product mean of operator concave functions}, we therefore obtain $ E=S. $
    \end{proof}

    This gives for $ k=1 $ the operator concavity in the positive half-axis of the function $ t\to t^p $
    for $ 0\le p\le 1. $ For $ k=2 $ we obtain concavity in the first quadrant
    of the function $ (t,s)\to t^ps^q $ for non-negative exponents with sum
    $ p+q\le 1. $ This is essentially Lieb's concavity theorem, cf. also Ando \cite[Corollary 6.2]{kn:ando:1979}
    who gave a different proof. The method of considering convex sets of exponents to prove concavity of
    the map $ A\to A^p\otimes A^q $ appeared in the unpublished notes \cite[Theorem IV.3]{kn:Ando:1978} by Ando.
    The same technique also appeared in a study of operator monotone functions \cite{kn:pedersen:1972},
    and very recently in
    a study of Morozova-Chentsov functions \cite[Remark 2.4]{kn:hansen:2006:2}.

    \section{New operator concave functions}

     Let us henceforth consider the functions
    \begin{gather}\label{function of k variables}
    f(t_1,\dots,t_k)=\frac{t_1}{t_1+\mu_1}\cdots\frac{t_k}{t_k+\mu_k}\qquad t_1,\dots,t_k>0,
    \end{gather}
    where $ \mu_1,\dots,\mu_k> 0. $

    \begin{definition}\label{definition: domain of concavity}
    We define the domain $ D_k(\mu_1,\dots,\mu_k)\subset\mathbf R^k_+ $ (abbreviated $ D_k $
    when there is no confusion) as the set of
    $ k $-tuples $ (t_1,\dots,t_k)\in\mathbf R_+^k $ such that the matrix
    \begin{gather}\label{matrix defining the domain}
    A_k(t_1,\dots,t_k)=\left(\begin{array}{cccc}
    \displaystyle\frac{2t_1}{\mu_1}  & -1 & \cdots & -1\\[3ex]
    -1 & \displaystyle\frac{2t_2}{\mu_2}  & \cdots & -1\\[3ex]
    \vdots & \vdots & \ddots & \vdots\\[3ex]
    -1 & -1 & \cdots & \displaystyle\frac{2t_k}{\mu_k}
    \end{array}\right)
    \end{gather}
    is positive semi-definite.
    \end{definition}

    It readily follows from the above definition that $ D_k $ is a closed convex set,
    and that $ (ct_1,\dots,ct_k)\in D_k $ for $ (t_1,\dots,t_k)\in D_k $ and $ c\ge 1. $

    \begin{proposition}
    The function $ f $ defined in (\ref{function of k variables}) is concave in the
    convex domain $ D_k. $ Furthermore, any open convex set in $ \mathbf R^k_+ $ in which $ f $ is
    concave is already contained in $ D_k. $
    \end{proposition}

    \begin{proof}
    The Hessian matrix $ H_f(t_1,\dots,t_k) $ of $ f $ is given by
    \[
    f(t_1,\dots,t_k)\left(\begin{array}{ccc}
\displaystyle\frac{-2\mu_1}{t_1(t_1+\mu_1)^2} &\displaystyle\frac{\mu_1\mu_2}{t_1t_2(t_1+\mu_1)(t_2+\mu_2)} &\cdots\\[3ex]
\displaystyle\frac{\mu_2\mu_1}{t_2t_1(t_2+\mu_2)(t_1+\mu_1)} &\displaystyle\frac{-2\mu_2}{t_2(t_2+\mu_2)^2} &\cdots\\[3ex]
\vdots & \vdots & \ddots
    \end{array}\right).
    \]
    If we introduce the manifestly positive semi-definite matrix
    \[
    P(t_1,\dots,t_k)=f(t_1,\dots,t_k)\left(\frac{\mu_i\mu_j}{t_it_j(t_i+\mu_i)(t_j+\mu_j)}\right)_{i,j=1}^k
    \]
    then the Hessian can be written as the Hadamard product
    \[
    H_f(t_1,\dots,t_k)=-A_k(t_1,\dots,t_k)\circ P(t_1,\dots,t_k),
    \]
    and since a Hadamard product is a principal submatrix of the tensor product, it follows
    that $ H_f(t_1,\dots,t_k) $ is negative semi-definite in the domain $ D_k. $ It hence
    follows that $ f $ is concave in $ D_k. $ Even though $ P(t_1,\dots,t_k) $ is a rank
    one operator it has a Hadamard inverse
    \[
    P^{\circ-1}(t_1,\dots,t_k)=\frac{1}{f(t_1,\dots,t_k)}\left(\frac{t_it_j(t_i+\mu_i)(t_j+\mu_j)}{\mu_i\mu_j}\right)_{i,j=1}^k,
    \]
    which is manifestly positive semi-definite in every point $ (t_1,\dots,t_k)\in\mathbf R^k_+, $  thus
    \[
    A_k(t_1,\dots,t_k)=-H_f(t_1,\dots,t_k)\circ P^{\circ-1}(t_1,\dots,t_k).
    \]
    If the Hessian were negative semi-definite in a point $ (t_1,\dots,t_k)\in\mathbf R^k_+ $ outside of $ D_k $ it would then
    follow that also $ A_k(t_1,\dots,t_k) $ is positive semi-definite, and this contradicts the definition
    of $ D_k. $ Therefore $ f $ is not concave in any open convex set outside of $ D_k. $
    \end{proof}

    We have shown that the function $ f $ defined in (\ref{function of k variables}) is concave in the
    domain $ D_k $ and nowhere concave outside of this domain. We will prove that
    $ f $ is in fact also operator concave in $ D_k, $ but first we need some preliminaries.

    \subsection{Generalized Hessian matrices}

    Matrix or operator convexity of a function of one or several variables may be
    inferred by calculating the so called generalized Hessian matrices
    \cite{kn:hansen:1997:2}. The theory is based on the structure theorem\footnote{In the reference we
    only considered functions defined in a product of open intervals, but the structure theorem is
    valid for functions defined in arbitrary open sets in $ \mathbf R^k. $}
    for the second Fréchet differential of the corresponding matrix function.

    Let $ f\colon D\to\mathbf R $ be a continuous function defined in an open set $ D\subseteq\mathbf R^k. $
    We say that a $ k $-tuple of bounded self-adjoint operators $ (x_1,\dots,x_k) $ acting on Hilbert spaces
    $ H_1,\dots,H_k $ is contained in the domain of $ f, $ if the
    product of the spectra $ \sigma(x_1)\times\cdots\times\sigma(x_k) $ is contained in $ D. $
    We may then proceed as in (\ref{functional calculus for operators}) to define the bounded
    self-adjoint operator $ f(x_1,\dots,x_k) $ acting on the tensor product $ H_1\otimes\cdots\otimes H_k. $

    A data set $ \Lambda $ for $ f $ of order $ (n_1,\dots,n_k) $ is a set of points
    in the domain $ D $ written on the form
    \begin{gather}\label{data set}
    \Lambda=\{(\lambda_{m_1}(1),\dots,\lambda_{m_k}(k))\in D\mid m_i=1,\dots,n_i\;\text{for}\; i=1,\dots,k\}.
    \end{gather}
    It may naturally be constructed from the eigenvalues of a $ k $-tuple of Hermitian
    matrices $ (x_1,\dots,x_k) $ of order $ (n_1,\dots,n_k) $ in the domain of $ f. $

    Suppose now that $ f\colon D\to\mathbf R $ has continuous partial derivatives up to the second order.
    To a data set $ \Lambda $ for $ f $ of order $ (n_1,\dots,n_k) $ as in (\ref{data set})
    and a $ k $-tuple of natural numbers $ (m_1,\dots,m_k) $ such that
    $ m_i\le n_i $ for $ i=1,\dots,k, $ the generalized Hessian matrix $ H(m_1,\dots,m_k) $ is defined
    \cite[Definition 3.1]{kn:hansen:1997:2} as the block matrix
    \[
    H(m_1,\dots,m_k)=
    \begin{pmatrix}
    H_{11}(m_1,\dots,m_k) & \cdots & H_{1k}(m_1,\dots,m_k)\\
    \vdots & \ddots & \vdots\\
    H_{k1}(m_1,\dots,m_k) & \cdots & H_{kk}(m_1,\dots,m_k)
    \end{pmatrix},
    \]
    where for $ u\ne s $ the $ n_u\times n_s $ matrix
    \[
    \begin{array}{l}
    H_{us}(m_1,\dots,m_k)=\\[1ex]
    \Bigl([\lambda_{m_1}(1)| \cdots |\lambda_{m_s}(s),\lambda_j(s)|\cdots|\lambda_p(u),\lambda_{m_u}(u)|
    \cdots|\lambda_{m_k}(k)]_f\Bigr)_{p,j}
    \end{array}
    \]
    while the $ n_s\times n_s $ matrix
    \[
    \begin{array}{l}
    H_{ss}(m_1,\dots,m_k)=\\[1ex]
    \Bigl(2[\lambda_{m_1}(1)| \cdots |\lambda_{m_s}(s),\lambda_p(s),\lambda_j(s)|\cdots
    |\lambda_{m_k}(k)]_f\Bigr)_{p,j}
    \end{array}
    \]
    for $ s=1,\dots,k. $
    The entries are second order partial divided differences of $ f $
    (the notation does not imply any particular order of the entries).
    Note that each generalized Hessian matrix
    is a quadratic and real symmetric matrix of order $ n_1+\cdots+n_k. $

    \begin{theorem}[The second Fréchet differential]
    Let $ f\colon D\to\mathbf R $ be a real $ p>2+k/2 $ times continuously differentiable
    function defined in an open set $ D\subseteq\mathbf R^k. $ Then the operator function
    \[
    (x_1,\dots,x_n)\to f(x_1,\dots,x_k),
    \]
    defined in $ k $-tuples $ (x_1,\dots,x_k) $ of bounded self-adjoint operators in the
    domain of $ f, $ is twice Fréchet differentiable. If this function is restricted to
    $ k $-tuples of Hermitian matrices $ (x_1,\dots,x_k) $ of order $ (n_1,\dots,n_k) $
    in the domain of $ f, $ then
    the expectation value of the second Fréchet differential can be written on the form
    \[
    \begin{array}{l}
    (d^2f(x)(h,h)\varphi\mid\varphi)\\[1ex]
    =\displaystyle\sum_{m_1=1}^{n_1}\cdots\sum_{m_k=1}^{n_k}
    \Bigl(H(m_1,\dots,m_k)\Phi^h(m_1,\dots,m_k)\mid
    \Phi^h(m_1,\dots,m_k)\Biggr),
    \end{array}
    \]
    where $ H(m_1,\dots,m_k) $ is a generalized Hessian matrix associated with $ f $ and the data
    set $ \Lambda $ constructed from
    the eigenvalues of the matrices $ (x_1,\dots,x_k). $ The vectors $ \Phi^h(m_1,\dots,m_k) $ are given by
    \[
    \Phi^h(m_1,\dots,m_k)=\left(\begin{array}{c}
                \Phi_1^h(m_1,\dots,m_k)\\
                \vdots\\
                \Phi_k^h(m_1,\dots,m_k)
                \end{array}\right),
    \]
    the $ k $-tuple of Hermitian matrices $ h=(h^1,\dots,h^k) $ is arbitrary but of order $ (n_1,\dots,n_k) $
    and the vectors
    \[
    \Phi_s^h(m_1,\dots,m_k)_{j_s}=h_{m_s j_s}^s
    \varphi(m_1,\dots,m_{s-1},j_s,m_{s+1},\dots,m_k)
    \]
    for $ j_s=1,\dots,n_s $ and $ s=1,\dots,k, $ and the tensor
    \[
    \varphi=\sum_{m_1=1}^{n_1}\cdots\sum_{m_k=1}^{n_k}
    \varphi(m_1,\dots,m_k)
    e_{m_1}^1\otimes\cdots\otimes e_{m_k}^k
    \]
    is expressed in terms of orthonormal bases of eigenvectors $ (e_{m_i}^i)_{m_i=1,\dots,n_i} $
    of each Hermitian matrix $ x_i $ in the $ k $-tuple $ (x_1,\dots,x_k). $
    \end{theorem}

    The form of the second Fréchet differential implies \cite[Exercises 3.1.8 and 3.6.4]{kn:flett:1980}
    the following result:

    \begin{corollary}\label{corollary: convexity of operator function}
     A real $ p>2+k/2 $ times continuously differentiable
    function  $ f\colon D\to\mathbf R $ defined in an open convex set $ D\subseteq\mathbf R^k $
    is matrix convex of order $ (n_1,\dots,n_k), $
    if to each data set $ \Lambda $ for $ f $ of order $ (n_1,\dots,n_k) $ all of the
    generalized Hessian matrices $ H(m_1,\dots,m_k) $ are positive semi-definite.
    \end{corollary}

    \begin{theorem}
    Let $ \mu_1,\dots,\mu_k>0 $ be positive real constants. The function
    \[
    f(t_1,\dots,t_k)=\frac{t_1}{t_1+\mu_1}\cdots\frac{t_k}{t_k+\mu_k}
    \]
    is operator concave in the domain $ D_k(\mu_1,\dots,\mu_k). $
    \end{theorem}

    \begin{proof}
    It is sufficient to prove that $ f $ is matrix concave of arbitrary order
    $ (n_1,\dots,n_k). $ For this purpose we consider an arbitrary data set $ \Lambda $ for $ f $
    of order $ (n_1,\dots,n_k) $ written as in (\ref{data set}). The multiplicative form of the function makes it
    simple to calculate the generalized Hessian matrices. We introduce the vectors
    \[
    a(i)=
    \left(\frac{\mu_i}{\lambda_1(i)+\mu_i},\dots, \frac{\mu_i}{\lambda_{n_i}(i)+\mu_i}\right)
    \in{\mathbf R}^{n_i}
    \]
    for $ i=1,\dots,k $ and calculate for $ u\ne s $ the entries
    \[
    \begin{array}{l}
    [\lambda_{m_1}(1)| \cdots
    |\lambda_{m_s}(s),\lambda_{j_s}(s)|\cdots|\lambda_{p_u}(u),\lambda_{m_u}(u)|\cdots|\lambda_{m_k}(k)]_f\\[2ex]
    =\displaystyle\frac{\lambda_{m_1}(1)}{\lambda_{m_1}(1)+\mu_1}\cdots\frac{\mu_s}{(\lambda_{m_s}(s)+\mu_s)
    (\lambda_{j_s}(s)+\mu_s)}\cdots\\[2ex]
    \displaystyle\hskip 10em\cdots\frac{\mu_u}{(\lambda_{p_u}(u)+\mu_u)(\lambda_{m_u}(u)+\mu_u)}\cdots
    \frac{\lambda_{m_k}(k)}{\lambda_{m_k}(k)+\mu_k}\\[3ex]
    =\displaystyle\frac{f(\lambda_{m_1}(1),\dots,\lambda_{m_k}(k))}{\lambda_{m_s}(s)\lambda_{m_u}(u)} a(u)_{p_u} a(s)_{j_s}
    \end{array}
    \]
    hence the block
    \[
    H_{us}(m_1,\dots,m_k)=\frac{f(\lambda_{m_1}(1),\dots,\lambda_{m_k}(k))}{\lambda_{m_s}(s)\lambda_{m_u}(u)}a(u)^t a(s).
    \]
    Similarly, we calculate the entries in the diagonal blocks
    \[
    \begin{array}{l}
    \displaystyle 2[\lambda_{m_1}(1)| \cdots |\lambda_{m_s}(s),\lambda_{p_s}(s),\lambda_{j_s}(s)|\cdots
    |\lambda_{m_k}(k)]_f\\[2ex]
    \displaystyle=\frac{2\lambda_{m_1}(1)}{\lambda_{m_1}(1)+\mu_1}\cdots
    \frac{-\mu_s}{(\lambda_{m_s}(s)+\mu_s)(\lambda_{p_s}(s)+\mu_s)(\lambda_{j_s}(s)+\mu_s)}
    \cdots\frac{\lambda_{m_k}(k)}{\lambda_{m_k}(k)+\mu_k}\\[4ex]
    \displaystyle=-2\frac{f(\lambda_{m_1}(1),\dots,\lambda_{m_k}(k))}{\mu_s\lambda_{m_s}(s)} a(s)_{p_s} a(s)_{j_s}
    \end{array}
    \]
    hence the block
    \[
    H_{ss}(m_1,\dots,m_k)=-2\frac{f(\lambda_{m_1}(1),\dots,\lambda_{m_k}(k))}{\mu_s\lambda_{m_s}(s)}a(s)^t a(s).
    \]
    In conclusion, the generalized Hessian matrices $ H(m_1,\dots,m_k) $ associated with the
    function (\ref{function of k variables}) and the data set (\ref{data set}) can be written on the form
    \[
    f(\lambda_{m_1},\dots,\lambda_{m_k})
    \left(\begin{array}{cccc}
    \displaystyle\frac{-2 a(1)^t a(1)}{\mu_1\lambda_{m_1}(1)}  &
    \displaystyle\frac{a(1)^ta(2)}{\lambda_{m_1}(1)\lambda_{m_2}(2)} &
    \cdots &
    \displaystyle\frac{a(1)^ta(k)}{\lambda_{m_1}(1)\lambda_{m_k}(k)}\\[3ex]
    \displaystyle\frac{a(2)^ta(1)}{\lambda_{m_2}(2)\lambda_{m_1}(1)} &
    \displaystyle\frac{-2 a(2)^t a(2)}{\mu_2\lambda_{m_2}(2)}  &
    \cdots &
    \displaystyle\frac{a(2)^ta(k)}{\lambda_{m_2}(2)\lambda_{m_k}(k)} \\[3ex]
    \vdots & \vdots & \ddots & \vdots\\[3ex]
    \displaystyle\frac{a(k)^ta(1)}{\lambda_{m_k}(k)\lambda_{m_1}(1)} &
    \displaystyle\frac{a(k)^ta(2)}{\lambda_{m_k}(k)\lambda_{m_2}(2)} &
    \cdots &
    \displaystyle\frac{-2 a(k)^t a(k)}{\mu_k\lambda_{m_k}(k)}
    \end{array}\right)
    \]
    where $ a(i)^t $ denotes the transpose of $ a(i). $ It can be written as the Hadamard
    product of the manifestly positive semi-definite block matrix
     \[
    f(\lambda_{m_1},\dots,\lambda_{m_k})
    \left(\begin{array}{cccc}
    \displaystyle\frac{a(1)^t a(1)}{\lambda_{m_1}(1)^2}  &
    \displaystyle\frac{a(1)^ta(2)}{\lambda_{m_1}(1)\lambda_{m_2}(2)} &
    \cdots &
    \displaystyle\frac{a(1)^ta(k)}{\lambda_{m_1}(1)\lambda_{m_k}(k)}\\[3ex]
    \displaystyle\frac{a(2)^ta(1)}{\lambda_{m_2}(2)\lambda_{m_1}(1)} &
    \displaystyle\frac{a(2)^t a(2)}{\lambda_{m_2}(2)^2}  &
    \cdots &
    \displaystyle\frac{a(2)^ta(k)}{\lambda_{m_2}(2)\lambda_{m_k}(k)} \\[3ex]
    \vdots & \vdots & \ddots & \vdots\\[3ex]
    \displaystyle\frac{a(k)^ta(1)}{\lambda_{m_k}(k)\lambda_{m_1}(1)} &
    \displaystyle\frac{a(k)^ta(2)}{\lambda_{m_k}(k)\lambda_{m_2}(2)} &
    \cdots &
    \displaystyle\frac{a(k)^t a(k)}{\lambda_{m_k}(k)^2}
    \end{array}\right)
    \]
    and the matrix
    $ -A_k(\lambda_{m_1}(1),\dots,\lambda_{m_k}(k)) $ defined in (\ref{matrix defining the domain}).

    All of the generalized Hessian matrices associated with $ f $ and $ \Lambda $ are thus negative semi-definite,
    hence it follows from Corollary~\ref{corollary: convexity of operator function} that $ f $ is matrix concave of order
    $ (n_1,\dots,n_k), $ and since this order is arbitrary, we conclude that $ f $ is operator concave.
    \end{proof}

    Since the above function $ f $ is operator concave in the largest domain in which it is concave, we realize
    that the associated generalized Hessian matrices of a certain order $ (n_1,\dots,n_k) $
    are negative semi-definite, if and only if $ f $ is matrix concave of the same order.
    This is in line with the conjecture (known to be true for functions of one variable)
    that positive semi-definiteness of the generalized Hessian matrices are not
    only sufficient but also necessary conditions for matrix convexity.

    \begin{corollary}
    Let $ \mu_1 $ and $ \mu_2 $ be positive real numbers, and let $ K $ be a Hilbert Schmidt operator.
    The operator function
    \[
    (A,B)\to\tr\left[\frac{A}{A+\mu_1}K^*\frac{B}{B+\mu_2}K\right],
    \]
    defined in pairs $ (A,B) $ of positive definite operators, is concave in the convex domain
    \[
    D_2(\mu_1,\mu_2)=\{(t_1,t_2)\in\mathbf R^2_+\mid t_1t_2\ge\mu_1\mu_2/4\}.
    \]
    \end{corollary}

    Note that the operator function in the corollary, for non-vanishing $ K, $ is not concave in any open
    convex set outside of $ D_2(\mu_1,\mu_2), $ not even its restriction to pairs of positive real numbers.

    \section{Appendix}

    \begin{theorem}
    The function
    \[
    f(t_1,\dots,t_k)=\frac{1}{t_1\cdots t_k}
    \]
    is operator convex in $ \mathbf R_+^k. $
    \end{theorem}

    \begin{proof}
    Let $ \Lambda $ be a data set for $ f $ of order $ (n_1,\dots,n_k) $  as in (\ref{data set}) and set
    \[
    a(i)=\left(\frac{1}{\lambda_1(i)},\dots,\frac{1}{\lambda_{n_i}(i)}\right)\in\mathbf R_+^{n_i}\qquad i=1,\dots,k.
    \]
    It is easy to calculate the generalized Hessian $ H(m_1,\dots,m_k) $ as
    \[
    f(\lambda_{m_1},\dots,\lambda_{m_k})
    \left(\begin{array}{cccc}
    2 a(1)^t a(1) & a(1)^ta(2)    & \cdots & a(1)^ta(k)\\[1ex]
    a(2)^ta(1)    & 2 a(2)^t a(2) & \cdots & a(2)^ta(k)\\[1ex]
    \vdots & \vdots & \ddots & \vdots\\[1ex]
    a(k)^ta(1)    & a(k)^ta(2)    & \cdots & 2 a(k)^t a(k)
    \end{array}\right)
    \]
    for any $ k $-tuple $ (m_1,\dots,m_k)\le(n_1,\dots,n_k). $ Since this matrix is manifestly positive semi-definite
    the assertion follows from Corollary \ref{corollary: convexity of operator function}.
    \end{proof}

    The above Theorem is due to Ando \cite[Theorem 5]{kn:ando:1979} who gave a very different proof.
    For $ k=2 $ the result may be derived from
    \cite[Corollary 8.1]{kn:lieb:1973:1} by
    using the identification $ \Phi $ introduced in the introduction.
    The result is fitting since $ -f $ is operator monotone as a function of $ k $ variables,
    cf. \cite[Page 17]{kn:hansen:2003:1}.

    \begin{corollary}
    The function
    \[
    f(t_1,\dots,t_k)=\frac{1}{t_1^{p_1}\cdots t_k^{p_k}}
    \]
    is for arbitrary exponents $ p_1,\dots,p_k\in [0,1] $ operator convex in $ \mathbf R_+^k. $
    \end{corollary}

    Lieb proved \cite[Corollary 3.1]{kn:lieb:1973:1} convexity of the mapping
    \[
    (A,B,K)\to\int_0^\infty\tr\left[\frac{1}{A+u}K^*\frac{1}{B+u}K\right]\,du
    \]
    in $ B(H)_+\times B(H)_+\times B(H)_{\mbox{\tiny HS}}, $ cf. also \cite{kn:ruskai:2005, kn:ohya:1993}.
    It is a triviality that the constituent mappings
    \[
    (A,B,K)\to\tr\left[\frac{1}{A+u}K^*\frac{1}{B+u}K\right]\qquad u>0
    \]
    are not (jointly) convex in $ B(H)_+\times B(H)_+\times B(H)_{\mbox{\tiny HS}}. $ But they
    are, as noted above, (jointly) convex in the first two variables.

    \begin{proposition}\label{proposition: joint convexity of map}
    The mapping $ (A,\xi)\to (A^{-1}\xi\mid\xi) $ is (jointly)
    convex for positive invertible operators $ A $ on a Hilbert
    space $ H, $ and vectors $ \xi\in H. $
    \end{proposition}

    \begin{proof}
    Ando noted\footnote{Since Ando offered no proof, we sketch (\ref{characterization of the harmonic mean})
    in the case where $ C $ is
    chosen as the harmonic mean. Use the identity $ 2(A^{-1}+B^{-1})=2A^{1/2}(1+A^{1/2}B^{-1}A^{1/2})^{-1}A^{1/2} $
    and multiply the inequality from the left and from the right with a diagonal block matrix with
    $ A^{-1/2} $ in the diagonal. This transformation reduces
    (\ref{characterization of the harmonic mean}) to an inequality between commuting operators.}
    \cite[Page 208]{kn:ando:1979} that the harmonic mean $ 2(A^{-1}+B^{-1})^{-1} $
    of two positive invertible operators $ A $ and $ B $ on a Hilbert space $ H $
    can be characterized as the maximum of all Hermitian operators $ C $ for which
    \begin{gather}\label{characterization of the harmonic mean}
    \begin{pmatrix}
    C & C\\
    C & C
    \end{pmatrix}\le
    2\begin{pmatrix}
    A & 0\\
    0 & B
    \end{pmatrix}.
    \end{gather}
    Replacing $ A $ and $ B $ with their inverses and inserting the Harmonic mean
    $ 2(A+B)^{-1} $ of $ A^{-1} $ and $ B^{-1} $ for $ C, $ we obtain the inequality
    \begin{gather}\label{block matrix inequality for inverses}
    \begin{pmatrix}
    (A+B)^{-1} & (A+B)^{-1}\\
    (A+B)^{-1} & (A+B)^{-1}
    \end{pmatrix}\le
    \begin{pmatrix}
    A^{-1} & 0\\
    0 & B^{-1}
    \end{pmatrix}
    \end{gather}
    which evaluated in block vectors $ (\xi, \eta) $ for $ \xi,\eta\in H $ may be written as
    \[
    \left(\left(\frac{A+B}{2}\right)^{-1}\left(\frac{\xi+\eta}{2}\right)\mid\left(\frac{\xi+\eta}{2}\right)\right)
    \le \frac{1}{2}\Bigl( (A^{-1}\xi\mid\xi) + (B^{-1}\eta\mid\eta)\Bigr).
    \]
    But this inequality is the desired result.
    \end{proof}

    The mapping $ A\to A\otimes B $ is linear for a fixed $ B, $ thus the mapping
    \[
    (A,\xi)\to ((A^{-1}\otimes B^{-1})\xi\mid\xi)
    \]
    is (jointly) convex for positive invertible operators $ A $ and $ B $ on a Hilbert space $ H $ and
    vectors $ \xi\in H\otimes H. $ By using the unitary map $ \Phi\colon H\otimes\bar H\to B(H) $
    introduced in the introduction, we obtain:

    \begin{proposition}
    The mapping
    \[
    (A,B,K)\to\tr\left[\frac{1}{A+u}K^*\frac{1}{B+v}K\right]\qquad u,v>0
    \]
    defined in $ B(H)_+\times B(H)_+\times B(H)_{\mbox{\tiny HS}} $ is (jointly)
    convex in any two of the three variables.
    \end{proposition}

    The joint convexity in say $ (A,K) $ may also be derived directly from the Lieb-Ruskai convexity
    theorem \cite[Remark after Theorem 1]{kn:lieb:1974} stating that the mapping $ (A,K)\to K^*A^{-1}K $
    is convex, where $ A $ is positive definite and invertible, and $ K $ is arbitrary.

    \begin{remark}\rm
     Lieb pointed out
    that Proposition~\ref{proposition: joint convexity of map} may be obtained also as a direct consequence of the
    Lieb-Ruskai theorem in the following way: Let $ B_\xi $ for an arbitrary vector $ \xi $ be defined as the operator
    $ B_\xi u=(u\mid v)\xi $ where $ v $ is a fixed unit vector. The mapping $ \xi\to B_\xi $ is linear, so the
    composed mapping  $ (A,\xi)\to B_\xi^* A^{-1} B_\xi $ is jointly convex.
    The desired result now follows by taking the expectation value in the vector $ v. $
    \end{remark}

    \begin{remark}\rm
    One may ask for which functions $ f $ defined in $ \mathbf R_+ $ the mapping
    \[
    (A,\xi)\to (f(A)\xi\mid\xi)
    \]
    is (jointly) convex. Obviously $ f $ has to be operator convex, and it follows immediately from
    Proposition~\ref{proposition: joint convexity of map} that any function of the form
    \begin{gather}\label{operator monotone decreasing functions}
    f(t)=\beta+\int_0^\infty\frac{1}{t+s}\,d\mu(s)\qquad \beta\in\mathbf R,
    \end{gather}
    where $ \mu $ is a positive measure with support in $ [0,\infty) $ such that the integrals
    $ \int (s^2+1)^{-1}\,d\mu(s) $ and $ \int s(s^2+1)^{-1}\,d\mu(s)$ both are finite, has the property.
    The functions of the form
    (\ref{operator monotone decreasing functions}) coincide with the class of operator
    monotone decreasing functions defined in the positive half-axis and bounded from below
    \cite[Page 9]{kn:hansen:2006:1}. But not all operator convex functions have the property. If we set
    $ f(t)=t^2 $ and choose the projections
    \[
    A_1=\begin{pmatrix}
        0 & 0\\
        0 & 1
        \end{pmatrix}\quad\text{and}\quad
    A_2=\frac{1}{2}\begin{pmatrix}
        1  & -1\\
        -1 & 1
        \end{pmatrix}
    \]
    together with the vectors $ \xi_1=(1,0) $ and $ \xi_2=(0,-1), $ then the difference
    \[
    \frac{(A_1^2\xi_1\mid\xi_1)+(A_2^2\xi_2\mid\xi_2)}{2}-
    \left(\left(\frac{A_1+A_2}{2}\right)^2\left(\frac{\xi_1+\xi_2}{2}\right)\mid\frac{\xi_1+\xi_2}{2}\right)
    =-\frac{1}{16}
    \]
    is negative, and this remains so if we perturb $ A_1 $ and $ A_2 $ slightly such that they become
    strictly positive.
    \end{remark}

    {\footnotesize


\begin{thebibliography}{10}

\bibitem{kn:Ando:1978}
T.~Ando.
\newblock Topics on \uppercase{O}perator \uppercase{I}nequalities.
\newblock Sapporo, 1978.
\newblock Unpublished notes.

\bibitem{kn:ando:1979}
T.~Ando.
\newblock Concavity of certain maps of positive definite matrices and
  applications to \uppercase{H}adamard products.
\newblock {\em Linear Algebra Appl.}, 26:203--241, 1979.

\bibitem{kn:bendat:1955}
J.~Bendat and S.~Sherman.
\newblock Monotone and convex operator functions.
\newblock {\em Trans. Amer. Math. Soc.}, 79:58--71, 1955.

\bibitem{kn:bhatia:1997}
R.~Bhatia.
\newblock {\em Matrix analysis}.
\newblock Springer, New York, 1997.

\bibitem{kn:birman:1972}
M.Sh. Birman and M.Z. Solomyak.
\newblock Notes on the function of spectral shift.
\newblock {\em Zap. Nauchn. Semin. LOMI}, 27:33--46, 1972.
\newblock (Russian).

\bibitem{kn:birman:1973}
M.Sh. Birman and M.Z. Solomyak.
\newblock Double \uppercase{S}tieltjes operator integrals, \uppercase{III}.
\newblock {\em Problems of Math. Physics, no. 6}, pages 27--53, 1973.
\newblock (Russian).

\bibitem{kn:epstein:1973}
H.~Epstein.
\newblock Remarks on two theorems of \uppercase{E}. \uppercase{L}ieb.
\newblock {\em Comm. Math. Phys.}, 31:317--325, 1973.

\bibitem{kn:flett:1980}
T.M. Flett.
\newblock {\em Differential Analysis}.
\newblock {Cambridge University Press}, {Cambridge}, 1980.

\bibitem{kn:hansen:1997:2}
F.~Hansen.
\newblock Operator convex functions of several variables.
\newblock {\em Publ. RIMS, Kyoto Univ.}, 33:443--463, 1997.

\bibitem{kn:hansen:2000:2}
F.~Hansen.
\newblock Operator inequalities associated with \uppercase{J}ensen's
  inequality.
\newblock In T.M. Rassias, editor, {\em Survey on Classical Inequalities},
  pages 67--98. Kluwer Academic Publishers, 2000.

\bibitem{kn:hansen:2003:1}
F.~Hansen.
\newblock Operator monotone functions of several variables.
\newblock {\em Math. Ineq. Appl.}, 6:1--17, 2003.

\bibitem{kn:hansen:2006:2}
F.~Hansen.
\newblock Characterization of symmetric monotone metrics on the the state space
  of quantum systems.
\newblock {\em arXiv:math-ph/0601056 v3}, pages 1--12, 2006.
\newblock To appear in Quantum Information and Computation.

\bibitem{kn:hansen:2006:1}
F.~Hansen.
\newblock Trace functions as \uppercase{L}aplace transforms.
\newblock {\em Journal of Mathematical Physics}, 47:043504 (2006).

\bibitem{kn:hansen:2005:3}
F.~Hansen and J.~Tomiyama.
\newblock Differential analysis of matrix convex functions.
\newblock {\em arXiv:math.OA/0601290 v1}, pages 1--17, 2006.
\newblock To appear in Linear Algebra Appl.

\bibitem{kn:hiai:2003:1}
F.~Hiai and H.~Kosaki.
\newblock {\em Means of \uppercase{H}ilbert space operators}.
\newblock {Lecture Notes in Mathematics}. {Springer}, Berlin, 2003.

\bibitem{kn:koranyi:1961}
A.~Kor{\'a}nyi.
\newblock On some classes of analytic functions of several variables.
\newblock {\em Trans Amer. Math. Soc.}, 101:520--554, 1961.

\bibitem{kn:kubo:1980}
F.~Kubo and T.~Ando.
\newblock Means of positive linear operators.
\newblock {\em Math. Ann.}, 246:205--224, 1980.

\bibitem{kn:lieb:1973:1}
E.~Lieb.
\newblock Convex trace functions and the
  \uppercase{W}igner-\uppercase{Y}anase-\uppercase{D}yson conjecture.
\newblock {\em Advances in Math.}, 11:267--288, 1973.

\bibitem{kn:lieb:1974}
E.H. Lieb and M.B. Ruskai.
\newblock Some operator inequalities of the \uppercase{S}chwarz type.
\newblock {\em Adv. in Math.}, 12:269--273, 1974.

\bibitem{kn:norlund:1924}
N.E. Nörlund.
\newblock {\em Vorlesungen \uppercase{\"U}ber \uppercase{D}ifferenzenrechnung}.
\newblock Springer Verlag, Berlin, 1924.

\bibitem{kn:ohya:1993}
M.~Ohya and D.~Petz.
\newblock {\em Quantum Entropy and its Use}.
\newblock {Springer Verlag}, Heidelberg, 1993.

\bibitem{kn:pedersen:1972}
G.K. Pedersen.
\newblock Some operator monotone functions.
\newblock {\em Proc. Amer. Math. Soc.}, 36:309--310, 1972.

\bibitem{kn:pusz:1975}
W.~Pusz and S.L. Woronowicz.
\newblock Functional calculus for sesquilinear forms and the purification map.
\newblock {\em Rep. Math. Phys.}, 8:159--170, 1975.

\bibitem{kn:ruskai:2005}
M.B. Ruskai.
\newblock Lieb's simple proof of concavity of $
  (\uppercase{A},\uppercase{B})\to
  \text{Tr\,}\uppercase{A}^p\uppercase{K}^\dag\uppercase{B}^{1-p}\uppercase{K}
  $ and remarks on related inequalities.
\newblock {\em arXiv:quant-ph/0404126 v3}, pages 1--14, 2005.

\end{thebibliography}


      \vfill

      {\noindent Frank Hansen: Department of Economics, University
       of Copenhagen, Studiestraede 6, DK-1455 Copenhagen K, Denmark.}
       }

      \end{document}